\documentclass[12pt]{article}
\usepackage{epsfig}
\usepackage{amsmath}
\usepackage{amssymb}
\usepackage{amsfonts}
\usepackage{latexsym}
\usepackage{color}

\def\be{\begin{equation}}
\def\ee{\end{equation}}
\def\ba{\begin{array}{c}}
\def\ea{\end{array}}

\def\ben{$$}
\def\een{$$}

\newcommand{\kt}{\rangle}
\newcommand{\br}{\langle}
\begin{document}

\titlepage

 \begin{center}{\Large \bf

Scattering along a complex loop

in a solvable ${\cal PT}-$symmetric model

 }\end{center}

\vspace{5mm}

 \begin{center}
Miloslav Znojil

\vspace{3mm}

Nuclear Physics Institute ASCR, 250 68 \v{R}e\v{z}, Czech
Republic\footnote{ e-mail: znojil@ujf.cas.cz }

\end{center}

\vspace{5mm}

\section*{Abstract}

%
%Quantum knots:
%
%${\cal PT}-$symmetric potentialless bound states

A non-unitary version of quantum scattering is studied via an
exactly solvable toy model. The model is merely asymptotically local
since the smooth path of the coordinate $x$ is admitted complex in
the non-asymptotic domain. At any real angular-momentum-like
parameter $\ell=\nu-1/2$ the reflection $R(\nu)$ and transmission
$T(\nu)$ are shown to change with the winding number (i.e.,
topology) of the path. The points of unitarity appear related to the
points of existence of quantum-knot bound states.

 \vspace{9mm}

\noindent
 PACS 03.65.Ge, 11.10.Kk, 11.30.Na, 12.90.+b

\newpage

\section{The concept of  ${\cal PT}-$symmetric toboggans \label{ruzka}}

In the older physics literature the use of  ${\cal PT}-$symmetric
quantum Hamiltonians (i.e., of the operators $H$ with the nonlinear
symmetry property ${\cal PT}H=H{\cal PT}$ where  ${\cal P}$ denotes
parity while ${\cal T}$ is complex conjugation which mimics time
reversal) has been purely {\em formal} (cf., e.g., Remark 4 in
Ref.~\cite{BG} dated 1993). This notion just referred, implicitly or
explicitly, to the use of the mathematical concept of Krein spaces
(tractable, for our present purposes, just as the Hilbert spaces
endowed with an auxiliary pseudometric ${\cal P}$, cf.
\cite{Tretter}).

A thorough change of the {\em physical} paradigm has been inspired
by Bender et al (cf., e.g., his own extensive account \cite{Carl} of
the history) who conjectured that the acceptance of the
non-Hermitian ${\cal PT}-$symmetric quantum Hamiltonians may be
perceived as ``legal" and that it might, and does, thoroughly enrich
our understanding of Quantum Theory as well as of the range of its
applicability. Via a thorough analysis of the quasi-one-dimensional
benchmark-model Hamiltonians
 \be
 H^{(\delta)}= -\frac{d^2}{d{x} ^2} - ({\rm i}x)^{2+\delta},
 \ \ \ \delta \geq 0
 \label{bebe}
 \ee
these authors clarified, in particular, that in spite of the
manifest non-Hermiticity (in the ``friendly but false"
representation space $L^2(\mathbb{R})$), these operators {\em may}
generate the real, discrete and below bounded spectrum of (in
principle, observable) energies. They have shown that one must only
define these operators in an appropriate, ``standard", {\em ad hoc}
Hilbert space ${\cal H}^{(S)}$ (we use the terminology of
papers \cite{SIGMAi,SIGMA} where further details have been
also discussed).

Naturally, the resulting  ${\cal PT}-$symmetric version of quantum
theory exhibits a number of immanent limitations and
counterintuitive features. The most visible one may be illustrated
via the benchmark-model Hamiltonian $H^{(\delta)}$. Its very
definition  requires a strongly counterintuitive, $\delta-$dependent
construction of the physical Hilbert space ${\cal H}^{(S)}$. This
space is being chosen as a rather exotic linear space of the square
integrable functions $f(x)$ which are defined along certain very
specific complex curves $x=x^{(\delta)}(s), \, s \in
(-\infty,\infty)$. In order to keep the spectrum real, these
``unobservable coordinate" curves must necessarily be shaped as
certain {\em complex}, left-right symmetric, downwards-oriented
hyperbolas in general. In addition, the $|s| \to \infty$ asymptotes
of these Hamiltonian-dependent curves must tend to parallel the
negative imaginary axis at the sufficiently large exponents
$\delta$~\cite{Carl}.

Many traditional model-building concepts are put under serious
questionmark in this setting. First of all, the ``traditional"
probability-density interpretation of the wave functions $\psi(x))$
is lost. Our wave functions become defined along the above-mentioned
complex curve $x=x(s)$ of the ``would-be coordinate". Next, one must
also speak about a coordinate-dependent kinetic energy and/or mass
$m=m(s)$ which may be complex. {\it In extremis}, this mass may even
happen to acquire a {\em purely negative} real value again (cf.,
e.g., \cite{Coulomb}).

One is forced to change (or at least to modify) also the traditional
thinking about the possible link of the mathematical model to any
experimental setup. An unexpectedly successful fulfillment of such a
requirement has been achieved, fortunately, in many models where
just a bound-state spectrum is to be studied \cite{Carl}. In
particular, an amazingly successful illustration of the underlying
non-Hermitian-representation approach to bound state spectra may be
found, under the nickname of ``interacting boson models", in nuclear
physics \cite{Geyer}.

The similar persuasive success is still lacking in the applications
of the same philosophy to the scattering.
In this alternative dynamical regime, the persistence of a number of
very serious difficulties has been revealed and reported by Jones
\cite{Jones}. He noticed that as a consequence of the change of the
traditional paradigm there emerge serious conceptual open questions
in the very formulation of the setup of the scattering experiment.
One of the most unpleasant obstacles emerged, for example, from an
unexpected formal conflict between the ``natural" requirements of
the ${\cal PT}-$symmetry of the Hamiltonian $H = p^2+V$, of the {\em
local} nature of the force ``prepared" at the short distances
($V=V(x)$  and of the asymptotically free and causality-preserving
nature of the incoming and outcoming asymptotic waves which are
assumed measured at the large distances (i.e., mathematically
speaking, of the scattering asymptotic boundary conditions).

The net conclusion of the latter study (cf. also the additional
thorough analysis \cite{Jonesdva} of the survival of the
long-range-correlation puzzles in the ``correct" Hilbert space
${\cal H}^{(S)}$) was that the ${\cal PT}-$symmetric quantum
scattering should only be interpreted as an effective theory where
one assumes the explicit presence of certain ``sinks" and ``sources"
in the space.

In the latter scenario, the unitarity of the scattering ceases to be
guaranteed of course. In our subsequent study \cite{scatt} we
proposed to weaken or circumvent such a scepticism and to reinstall
the unitarity of the quantum scattering by ${\cal PT}-$symmetric
obstacles via the use of certain short-ranged nonlocalities in the
potentials.

In another approach reported in Ref.~\cite{toboscatt} we proposed to
try to move one step further. In place of using the rather
complicated non-local integral-operator kernels $V=V(x,x')$ we
decided to keep the potential local, $V=V(x)$, and to introduce a
new degree of freedom via a short-ranged ``space-smearing" term
attached to the mass ($m \to m(x)$) or, better, to the energy term
in the Schr\"{o}dinger equation ($E \to E \times W(x)$).
In spite of the presence of the new term $W \neq I$, the resulting
generalized Schr\"{o}dinger equations (or, in our terminology,
``Sturm-Schr\"{o}dinger equations" \cite{preHendrik}) still keep the
trace of the phenomenological ambitions of ${\cal PT}-$symmetric
models. On mathematical side, they proved also friendly and
tractable by the standard ``Hermitization" trick mediated,
constructively, by the transition to the suitable {\em ad hoc}
Hilbert space ${\cal H}^{(S)}$. This space has been shown to exist
and to remain amenable to constructive considerations
\cite{Hendrik}.

The climax of the story comes with the idea that the ${\cal
PT}-$symmetric ``Sturm-Schr\"{o}dinger" differential equations
(defined, say, along the real line of coordinates $x(s)\equiv
\mathbb{R}$) may be mathematically simplified via a suitable change
of variables. This makes them equivalent to the
``usual-Schr\"{o}dinger" differential equations defined,
anomalously, along the so called ``tobogganic" curve of ``would-be"
complex coordinates $x(s)=x^{(tobog.)}(s)\neq \mathbb{R}$.
Typically, the latter ``tobogganic" contour connects several Riemann
sheets of the wave function \cite{riem} so that the quantitative
analysis of its spectrum (or scattering) becomes complicated.

Our previous letter \cite{knots} described an extremely
elementary analytic (and, moreover, non-numerically solvable)
tobogganic model of bound states. In our present paper we just
intend to complement this study by a parallel description of the
same model in the scattering dynamical regime.
Firstly, we shall demand that our tobogganic ``would-be
coordinate" curves $x=x(s)$ remain {\em asymptotically real}
(i.e., asymptotically observable, with $x (s) \sim s \in \mathbb{R}$
at $|s| \gg 1$). Secondly, our ``over-schematic" toy-model potential
will be taken over from Ref.~\cite{knots}. In this manner, the exact
solvability of the related scattering ``Gedanken-experiment" will be
retained.

\section{Elementary model \label{soudruz}}

The ordinary linear differential equation
 \be
  -\frac{d^2}{d{x} ^2}\,\psi ({x})+ \frac{\ell(\ell+1)}{{x}^2}
  \,\psi ({x})+ V({x})
  \,\psi ({x})= E \,\psi ({x})\,
   \label{SEnotfree}
 \ee
is often encountered in the textbooks on quantum mechanics where it
emerges, with $x \in \mathbb{R}^+$, as the so called radial part of
the Schr\"{o}dinger equation in $D$ dimensions. Three years ago we
proposed~\cite{knots} an alternative quantum interpretation of
Eq.~(\ref{SEnotfree}) in which the path ${\cal C}={\cal C}^{(N)}$ of
the ``coordinate" $x$ has been allowed complex, forming a
loop-shaped curve, $N-$times encircling the branch point of $\psi
({x})$ at $x=0$.

For the sake of simplicity the external potential itself has been
assumed absent, $V({x})=0$. The resulting bound-state solutions
$\psi ({x})$ of Eq.~(\ref{SEnotfree}) were then given in closed form
called, due to its topological origin, ``quantum knot". It has been
emphasized that the ``quantum knot" solutions could find their
natural and consistent physical interpretation, e.g., within the
framework of the so called ${\cal PT}-$symmetric Quantum Mechanics
(cf., e.g., the recent reviews \cite{Carl,ali} for a more
detailed exposition of this formalism).

%
%
%The key motivation of our present letter can be traced back to the
%two papers by Jones \cite{Jones,Jonesdva}. He revealed and
%demonstrated that in the generic ${\cal PT}-$symmetric model the
%necessary loss of the observability of the ``coordinate" $x$ renders
%the original bound-state recipe inapplicable in the regime of
%scattering. In such a situation one can either modify the recipe
%(one of the eligible constructive illustrations of such an approach
%may be found in Ref.~\cite{scatt}) or work just with an
%``effective", non-unitarity version of the scattering theory (note
%that in Ref.~\cite{Coul} we demonstrated the applicability of the
%latter approach to Eq.~(\ref{SEnotfree}) with Coulombic $V(x) \sim
%1/x$).
%
%
%
%==========
%
%physics behind
%
%
%
%A non-unitary effective 1D quantum scattering exhibiting a
%long-range balance between its sinks and sources (a.k.a. ${\cal
%PT}-$symmetry) is studied via an exactly solvable toy model. The
%model is merely asymptotically local since the smooth path of the
%coordinate $x$ is admitted, in the non-asymptotic domain, complex.
%At any real angular-momentum-like kinematical parameter
%$\ell=\nu-1/2$ the reflection $R(\nu)$ and transmission $T(\nu)$ are
%shown nontrivial. Their variability (such that $z(\nu)=|R|^2+|T|^2
%\neq 1$ in general) is found controlled by the winding number (i.e.,
%topology) of the path. The points of unitarity (such that
%$z(\nu)=1$) appear related to the points of existence of
%quantum-knot bound states.

\begin{figure}[h]                     %instead of \begin{figure}[t]
\begin{center}                         %instead of \begin{center}
\epsfig{file=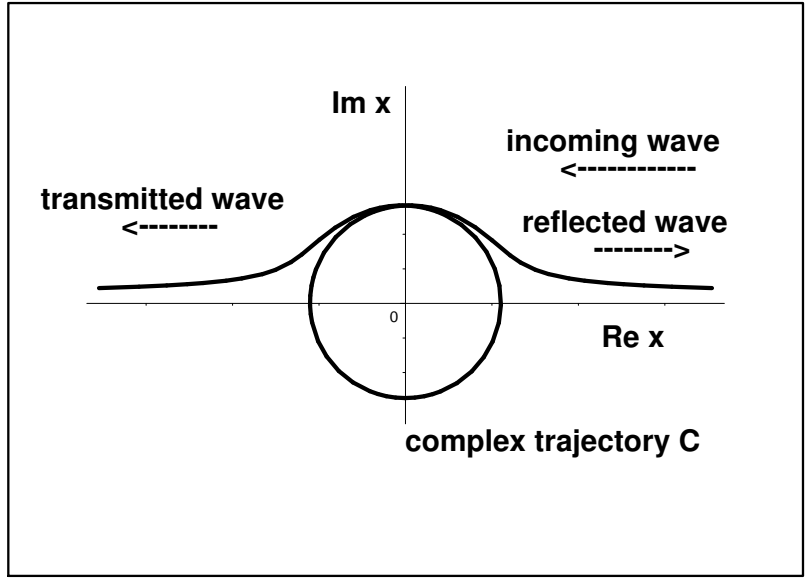,angle=270,width=0.7\textwidth}
\end{center}                         %instead of \end{center}
\vspace{-2mm} \caption{The arrangement of scattering along a
loop-shaped complex path which circumscribes the origin and which
coincides with the real line at large $|x|$.
 \label{fione}}
\end{figure}

In our present letter, as we already mentioned, we intend to
complement the results of paper \cite{knots} by a parallel study of
the problem of scattering. In a way paralleling and completing our
previous study \cite{knots} we shall outline a few most interesting
consequences of the acceptance of the ${\cal PT}-$symmetrization
strategy in the case of the vanishing potential, $V({x})=0$. The
presence of just a ``minimal" dynamical input will be compensated by
the topologically nontrivial choice of the loop-shaped paths
$x^{(tobog.)}(s)$ or rather ${\cal C}={\cal C}^{(N)}$ marked by a
winding number $N = 1, 2, \ldots$. A characteristic one-loop example
of such a path with $N=1$ is displayed in Fig.~\ref{fione}.

%\newpage
%********** Figure 1 zde
%\newpage

For the scattering-experiment arrangement as indicated in
Fig.~\ref{fione} we shall postulate that $x$ (= the argument of the
wave function $\psi(x)$) and $q$ (= the real eigenvalue of a
suitable particle-position operator $\hat{Q}$) will coincide for
large  $|q| \gg 1$. Moreover, we may follow Fig.~\ref{fione} and
specify the positive asymptotic domain of $x \approx q \gg 1 $ as
the domain of the incoming wave while the large and negative
positions  $x \approx q \ll -1 $ will be assigned to the outcoming,
transmitted wave. Thus,  we shall complement Eq.~(\ref{SEnotfree})
by the scattering boundary conditions
 \be
 \psi(x) =
 \left \{
 \begin{array}{cc}
 e^{-{\rm i}\kappa x} + R\,e^{{\rm i}\kappa x},& \ \ \ \ x \gg 1,\\
 T\,e^{-{\rm i}\kappa x} ,& \ \ \ \ x \ll -1
 \ea
 \right .
 \label{bcs}
 \ee
at any energy $E=\kappa^2$.

\section{Scattering dynamical regime \label{soudruzka}}

\subsection{The reflection and transmission coefficients}

The decisive advantage of our choice of $V(x)=0$ in Schr\"{o}dinger
Eq.~(\ref{SEnotfree}) is its exact solvability, at any energy
$E=\kappa^2$, in terms of Hankel functions~\cite{Ryzhik},
 \be
 \psi(x) = c_1\,\sqrt{x}\,H^{(1)}_\nu(\kappa\,x)
 +c_2\,\sqrt{x}\,H^{(2)}_\nu(\kappa\,x)\,,\ \ \ \  \nu =
 \ell+1/2\,, \ \ \ \ x \in {\cal C}^{(N)}\,.
 \label{ansat}
 \ee
The asymptotics of these solutions may easily be derived since at
$|{\rm arg}\,z| < \pi$ and ${\rm Re}\,\nu > -1/2$ we
have~\cite{Ryzhik}
 \ben
 \sqrt{\frac{\pi z}{2}}\,H^{(1)}_\nu(z)=
 \exp\left [{\rm i}\left (z-\frac{\pi(2\nu+1)}{4}
 \right )\right ]\,\left (1-\frac{\nu^2-1/4}{2{\rm i} z} + \ldots
 \right )
 \,,
 \een
 \ben
 \sqrt{\frac{\pi z}{2}}\,H^{(2)}_\nu(z)=
 \exp\left [-{\rm i}\left (z-\frac{\pi(2\nu+1)}{4}
 \right )\right ]\,\left (1+\frac{\nu^2-1/4}{2{\rm i} z} + \ldots
 \right )
 \,
 \een
In other words, we may abbreviate $\kappa\,x=z(x)=z$ and eliminate
 \ben
 \exp\left ({\rm i}\,\kappa\,x\right )=
 \sqrt{\frac{\pi \kappa\,x}{2}}\,H^{(1)}_\nu(\kappa\,x)\,
 \exp\left [-{\rm i}\left (\frac{\pi(2\nu+1)}{4}
 \right )\right ]\,\left (1-\frac{\nu^2-1/4}{2{\rm i} \kappa\,x} + \ldots
 \right )
 \,,
 \een
 \ben
 \exp\left (-{\rm i}\,\kappa\,x\right )=
 \sqrt{\frac{\pi \kappa\,x}{2}}\,H^{(2)}_\nu(\kappa\,x)\,
 \exp\left [{\rm i}\left (\frac{\pi(2\nu+1)}{4}
 \right )\right ]\,\left (1+\frac{\nu^2-1/4}{2{\rm i} \kappa\,x} + \ldots
 \right )
 \,.
 \een
These formulae may be inserted in boundary conditions (\ref{bcs})
yielding, in the leading order of approximation,  the following
closed formula for the ``far right" wave function  $\psi(x)\approx
 e^{-{\rm i}\kappa\,x} + R\,e^{{\rm i}\kappa\,x}\approx$
 \be
 \approx \sqrt{\frac{\pi \kappa\,x}{2}}\,H^{(2)}_\nu(\kappa\,x)\,
 \exp\left [{\rm i}\left (\frac{\pi(2\nu+1)}{4}
 \right )\right ]
 + R\,
 \sqrt{\frac{\pi \kappa\,x}{2}}\,H^{(1)}_\nu(\kappa\,x)\,
 \exp\left [-{\rm i}\left (\frac{\pi(2\nu+1)}{4}
 \right )\right ]\,
 \label{abcs}
 \ee
at $x \gg 1$, as well as the complementary asymptotic estimate of
the ``far left" wave function
 \be
 \psi(x) \approx
 T\,e^{-{\rm i}\kappa\,x}\approx T\,
\sqrt{\frac{\pi \kappa\,x}{2}}\,H^{(2)}_\nu(\kappa\,x)\,
 \exp\left [{\rm i}\left (\frac{\pi(2\nu+1)}{4}
 \right )\right ]\,,\ \ \ \ \ x \ll -1\,.
  \label{bbcs}
 \ee
The right-hand-side expression in formula (\ref{bbcs}) defines in
fact a particular {\em exact} solution of Eq.~(\ref{SEnotfree})
which may be analytically continued along the {\em whole} complex
integration path  ${\cal C}={\cal C}^{(N)}$. This path, by
construction, moves from the left infinity to the right infinity
while performing $N=-m/2 \geq 1$ clockwise rotations around the
origin. This means that at the points belonging to the ``far right"
part of the curve ${\cal C}^{(N)}$  our function becomes equal to
the expression
 \be
 T\,
 \sqrt{\frac{\pi z}{2}}\,H^{(2)}_\nu(ze^{{\rm i}m\pi})\,
 \exp\left [{\rm i}\left (\frac{\pi(2\nu+1)}{4}
 \right )\right ]\,\,,\ \ \ \ \ x \gg +1\,.
  \label{cbbcs}
 \ee
At the same time, such a function has to match the $x \gg 1$
boundary conditions (\ref{bcs}) or (\ref{abcs}). Thus, in a way used
in Ref.~\cite{knots} it is now sufficient to recall formula 8.476.7
of ref.~\cite{Ryzhik},
 \be
 H^{(2)}_\nu\left (ze^{{\rm i}m\pi}\right )=
 \frac{\sin (1+m)\pi\nu}{\sin \pi \nu}\,
 H^{(2)}_\nu(z)+
 e^{{\rm i}\pi\nu}\,
 \frac{\sin m\pi\nu}{\sin \pi \nu}\,
 H^{(1)}_\nu(z)\,
 \label{nuit}
 \ee
and to insert it in Eq.~(\ref{cbbcs}). By comparing the result with
Eq.~(\ref{abcs}) one arrives at our final explicit formulae
 \be
 R=R(\nu)=\frac{\sin m \pi \nu}{\sin (m+1) \pi \nu}\,e^{{\rm i}(4\nu+1)\pi/2}
 \ee
and
 \be
 T=T(\nu)=\frac{\sin  \pi \nu}{\sin (m+1) \pi \nu}\,
 \ee
which characterize the result of the scattering at any topological
dynamical-input parameter $N=-m/2= 0, 1, \ldots\,$.

\subsection{The points of unitarity vs. quantum knots}

\subsubsection{Single loop, $N=1$.}

In the one-loop arrangement of Fig.~\ref{fione} with $N=1$,
$T(\nu)=-1$ and $|R(\nu)|=2\cos \pi \nu$ one arrives at the
elementary formula
 \be
 z=z(\nu)=|T|^2+|R|^2=1+4\cos ^2 \pi \nu \geq 1\,.
 \ee
This shows that the unitarity of the scattering is solely being
preserved at the integer values of $\ell=\nu-1/2$ or, in other
words, just in the reflectionless cases. In the context of
Ref.~\cite{knots} it is remarkable to notice that the one-loop
quantum knots {\em also} proved to exist {\em precisely} at the same
values of $\ell$.

%\newpage
%********** Figure a2 zde
\begin{figure}[h]                     %instead of \begin{figure}[t]
\begin{center}                         %instead of \begin{center}
\epsfig{file=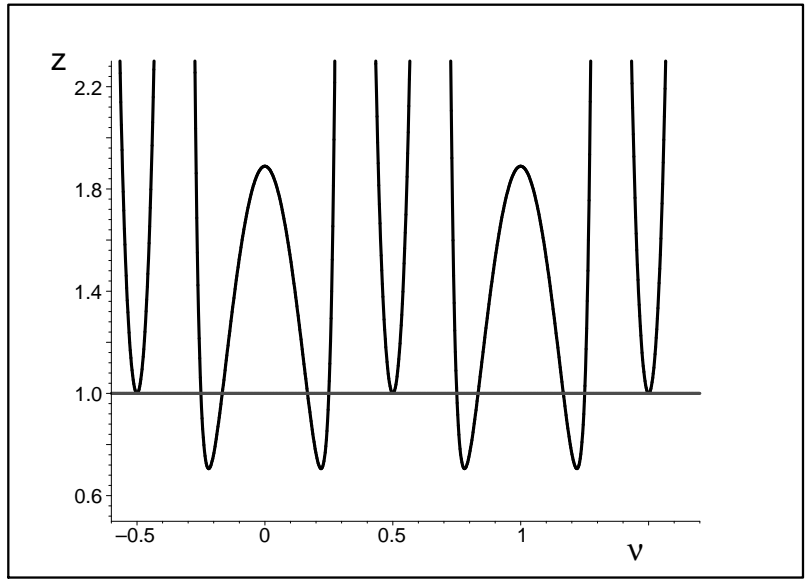,angle=270,width=0.7\textwidth}
\end{center}                         %instead of \end{center}
\vspace{-2mm} \caption{The values of $z=|T|^2+|R|^2$ and the $z=1$
points of unitarity at $N=2$.
 \label{ttro}}
\end{figure}
%\newpage

\subsubsection{Double loop, $N=2$.}

Starting from the next, two-loop scenario the coefficients $R$ and
$T$ cease to be bounded. Their trigonometric form remains elementary
but the violation of the unitarity acquires a more subtle form (cf.
Fig.~\ref{ttro}). In particular, there emerge closed intervals of
$\ell$ in which  $z=|T|^2+|R|^2\leq 1$ {\em and} in which the
reflection $R$ does not vanish.

The above-mentioned remarkable relationship between the incidental
$z=1$ unitarity of the scattering and the existence of the quantum
knots of Ref.~\cite{knots} merely {\em partially} survives in the
two-loop case. In the interval of $\nu \in (0,1)$, for example, one
finds as many as six roots of the unitarity constraint $z(\nu)=1$
(including multiplicity, viz., the values of $\nu = 1/6, 1/4, 1/2,
1/2, 3/4$ and $5/6$). Merely three of them (viz., the values of $\nu
=  1/4, 1/2$ and $3/4$) have been shown to imply the existence of a
two-loop quantum knot in Ref.~\cite{knots}.

%\newpage
%********** Figure a3 zde
\begin{figure}[h]                     %instead of \begin{figure}[t]
\begin{center}                         %instead of \begin{center}
\epsfig{file=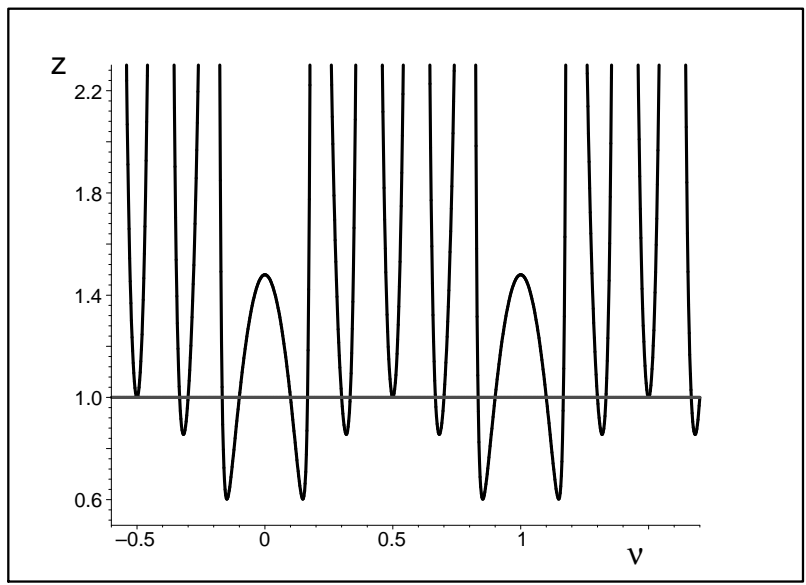,angle=270,width=0.7\textwidth}
\end{center}                         %instead of \end{center}
\vspace{-2mm} \caption{The values of $z=|T|^2+|R|^2$ and the $z=1$
points of unitarity at $N=3$.
 \label{firo}}
\end{figure}
%\newpage

\subsubsection{Multiple loops, $N>2$.}

At the higher winding numbers the overall pattern remains very
similar. In particular, the next, $N=3$ sample of the function
$z=z(\nu)=|T|^2+|R|^2$ is given in Fig.~\ref{firo}. The inspection
of this picture enables us to reveal that in the interval of $\nu
\in (0,1)$ there exist five values of $\nu$ guaranteeing the
existence of a three-loop quantum knot~\cite{knots} (viz., $\nu =
1/6,2/6,3/6,4/6$ and $5/6$) but as many as ten roots of equation
$z(\nu)=1$ (viz., the values of $\nu = 1/10,1/6, 3/10, 1/3,1/2, 1/2,
2/3,7/10, 5/6$ and $9/10$).

An extrapolation of this pattern to an arbitrary winding number
seems straightforward. In a remarkable and nontrivial manner it
interrelates the scattering and bound states in the manner which
resembles the well known correspondence between the bound states and
poles of analytic S-matrix.

\section{Discussion}

\subsection{General alocal quantum systems}

It is well known that the overall, abstract postulates of Quantum
Theory do {\em not} require the observability (i.e., reality) of the
one-dimensional coordinate $x$. The author of Ref.~\cite{Hoo}, for
example, emphasized that the quantity $x\in \mathbb{R}$ plays a
double role in quantum physics. For the time being let us call them
``physical" (A) and ``mathematical" (B). The former role (A) means
that whenever our coordinate $x$ appears as an argument in a wave
function, $\psi=\psi(x)$, we immediately -- and almost always
tacitly -- assume that, firstly, this wave function describes a
point particle moving along a straight line while, secondly, the
position of this point particle is measurable and represented by
such an operator $Q$ that $Q \psi(x) = x\,\psi(x)$.

It is important to notice that whenever one starts working with the
same-looking wave function  $\psi=\psi(x)$ in scenario (B), there
is, first of all, no implication concerning physics. The purely
formal reason is that all of the possible concrete realizations of
the abstract separable Hilbert space are mutually unitarily
equivalent. In this sense, a state of any quantum system may be
represented, if we so decide, by a quadratically integrable complex
function $\psi(x) \in \mathbb{L}^2(\mathbb{R}, d\mu)$. In general,
the quantum system in question need not even be assigned any
measurable coordinate at all.

In the latter case the meaning of the {\em real argument} of
$\psi(x)$ may remain {\em physical} but still {\em very different}
from a spatial position of a localized particle (cf. Ref.~\cite{Hoo}
for some most elementary illustrative examples). In an extreme
alternative, the argument $x$ of $\psi(x)$ (defined, in the Dirac's
notation, as equal to a bra-ket overlap, $\psi(x)=\br x|\psi\kt$)
may even be chosen {\em complex} (cf. \cite{Carl} for a nice recent
review of some interesting and promising merits of such an option).

\subsection{Locally alocal systems}

Once we decide to work with the entirely formal, Riesz-basis-related
concept of the overlaps $\br x|\psi\kt$ which are merely ``numbered"
by the quantities (or rather ``indices") $x$ forming, as in our
present paper, a left-right symmetric (often called ${\cal
PT}-$symmetric \cite{Carl}) complex curve ${\cal C}$, we may still
try to generate the time-evolution of the system via a sufficiently
simple {\em ad hoc} Hamiltonian operator $H$. One of the most
persuasive illustrative example of the appeal of such a
model-building direction (admitting that a measurable coordinate
does not exist at all) has been offered by Witten \cite{Witten}.
While he tried to understand and/or classify the possible mechanisms
of a breakdown of symmetry (called, nowadays, supersymmetry,
connecting fermions and bosons) he proposed an elementary
partitioned toy-model Hamiltonian
 \be
 H =\left (
 \begin{array}{cc}
 H^{(-)}&0\\
 0&H^{(+)}
 \end{array}
 \right )\,,\ \ \ \ \
 H^{(\pm)}= -\frac{d^2}{dx^2} + V^{(\pm)}(x)\,,
 \ \ \ \ \ x \in \mathbb{R}
 \ee
where $x$ {\em was not} and observable of course \cite{Khare}.

In a way reemphasized by many other authors~\cite{Jones,ali,Hoo} we
need not really insist on the observability of the coordinate $x$,
especially when we study bound states. The situation is less liberal
in the context of scattering in which one may truly appreciate
having the {\em physical} concept of the {\em real and measurable}
position $q\in \mathbb{R}^d$. In this context our present message is
that one can weaken the requirement and preserve the concept of the
locality {\em just} in the asymptotic spatial domains.

In this context we introduced here the idea of motion of a quantum
(quasi)particle along the path of coordinates of Fig.~\ref{fione}
which are only real (i.e., in principle, observable) asymptotically.
Naturally, such an assumption is {\em sufficient} for the imposition
of the ``realistic" asymptotic boundary conditions and for the
related constructions of certain effective ${\cal PT}-$symmetric
one-dimensional models of scattering by a local potential $V(x)$ as
studied in the recent literature \cite{Jones,Cannata}.

\subsection{Open questions}

In a way encouraged by the phenomenological as well as methodical
success of similar quantum models the key innovation as offered in
the present letter lies in the {\em topological} nontrivality of our
present, locally complex trajectories ${\cal C}^{(N)}$. Such an
extension of the perspective remains compatible with the current
expectations \cite{Cannata} that the reflection coefficient $R$ and
the transmission coefficient $T$ will not obey the unitarity
constraint $z=|T|^2+|R|^2 \neq 1$ in general. Naturally, this
freedom opens a number of possible physical interpretations ranging
from the theory of open systems up to the possible fructification of
the philosophy of the present elementary example in the context of
the path-integral quantization or, alternatively, in the recently
fashionable context of the models using the concept of a
coordinate-dependent mass.

In the language of physics the generic non-unitarity $z \neq 1$ may
be perceived as a natural consequence of the admitted presence of
certain ``sources" and ``sinks". These implicit non-Hermitian forces
may be intuitively expected to be rooted not only in the imaginary
part of the potential but also \cite{tobog} in the effects caused by
the topologically nontrivial deformations of the integration paths
as sampled by Eq.~(\ref{fione}). This type of connection has been
supported here by a rather unexpected observation that the standard
connection between bound states and poles of S-matrix might also
find its analogue in the present loop-shaped-path context.

In conclusion, one should emphasize that the loop-shape-generated
balance between sinks and sources as imposed by ${\cal PT}-$symmetry
is merely long-ranged and, hence, too weak for a reinstallation of
the unitarity. Although an overall explanation of this fact does not
require any particularly sophisticated mathematics \cite{scatt},
there still exist a few open questions which have been formulated by
{\em physicists}. Jones \cite{Jones}, in particular, imagined that
the model-building freedom offered by the {\em complex local}
potentials $V(x)$ is in fact rather expensive, {\em both} in terms
of the feasibility of the constructions {\em and} in terms of the
appropriate {\em preparation} of any external physical interaction
$V(x)$. In this sense, our present ``exceptional" choice of $V(x)=0$
offered one of particularly efficient ways of circumventing the
problem.

\section*{Acknowledgements}

Supported by the GA\v{C}R grant Nr. P203/10/1433, by the M\v{S}MT
``Doppler Institute" project Nr. LC06002 and by the NPI
Institutional Research Plan AV0Z10480505.

%\newpage

\end{document}